\documentclass{ws-procs10x7}

\begin{document}

\title{Semileptonic $D$ Decays from CLEO and BELLE}

\author{Yongsheng Gao}

\address{Physics Department, Southern Methodist University,
          Dallas, TX 75275, USA}

\twocolumn[\maketitle\abstract{Recent semileptonic $D$ meson decay
results from CLEO and BELLE are summarized, including the improved
measurements of absolute branching fractions for exclusive $D^{0}$
semileptonic decays into $K^{-}e^{+}\nu$, $\pi^{-}e^{+}\nu$
and $K^{*-}e^{+}\nu$, and the first observation and absolute branching
fraction measurement of $D^{0} \to  \rho^{-}e^{+}\nu$ with the first
CLEO-c data sample.
}]

\section{Introduction}

Semileptonic $D$ meson decays are of great physics interest because
their description is relatively simple. 
The decay matrix for semileptonic $D$ meson decay decouples into 
a weak current component (describing the $W \ell \nu_{\ell}$ vertex), 
and a strong current term (for the $W c \bar{q}$ vertex) that is 
parameterized through form factor
functions of the invariant mass ($q^{2}$) of the $W$ exchanged.
The form factors can not be easily computed in quantum 
chromodynamics (QCD) since they are affected by significant
nonperturbative contributions. 
That is the main source of uncertainty in the extraction of the 
CKM matrix element from the simple decay processes. 
Precise experimental measurements are needed to guide 
theoretical progress in this area. 
Charm semileptonic decays allow a measurement of the form factors
and CKM matrix elements $V_{cs}$ and $V_{cd}$.
Using Heavy Quark Effective Theory (HQET) or lattice gauge techniques, 
the measured charm form factor can be related to those needed to 
interpret $b \to u$ transition and the measurement of
the CKM matrix element $V_{ub}$.

\begin{figure}
\epsfxsize190pt
\figurebox{190pt}{220pt}{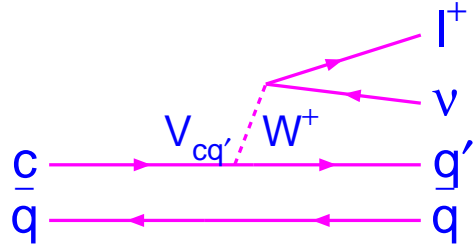}
\caption{Feynman diagram for charm meson semileptonic decays.}
\label{fig1}
\end{figure}

While measurements of form factors in all charm exclusive semileptonic
decays are important, those in pseudoscalar-to-pseudoscalar
transitions are the easiest to perform.
The differential decay rate for the exclusive semileptonic decay 
$D\to P\ell\nu$ ($P$ stands for a pseudoscalar meson) with 
the electron mass effects neglected can be expressed 
as~\cite{theory}:
\begin{eqnarray}
\frac{d\Gamma}{dq^2}=\frac{G_F^2}{24\pi^3}\left|V_{cq'}\right|^2p_P^3
\left|f_+(q^2)\right|^2
\label{diffrate} 
\end{eqnarray}   
where $G_F$ is the Fermi coupling constant, $q^2$ is the 
four-momentum transfer squared between the parent $D$ meson 
and the final state meson, $p_P$ is the momentum 
of the pseudoscalar meson in the $D$ rest frame, and $V_{cq'}$ is 
the relevant CKM matrix element, either $V_{cs}$ or $V_{cd}$. 
$f_+(q^2)$ is the form factor that measures the probability that 
the flavor changed quark $q'$  and the spectator quark $\bar q$ 
in Figure~\ref{fig1} will form a meson in the final state.  

Because the semileptonic $D$ meson decay results from BELLE were not available
by ICHEP04, I'll only describe the recent work and results on this topic from 
CLEO.

\section{$D^{0}\to \pi^{-}\ell^{+}\nu$ and
            $D^{0}\to   K^{-}\ell^{+}\nu$ from CLEO III}

We present a study of the decay $D^{0}\to \pi^{-}\ell^{+}\nu$
and $D^{0}\to   K^{-}\ell^{+}\nu$ where $\ell$ = $e$ or $\mu$
with $e^{+}e^{-} \to c\bar{c}$ events collected at and just
below the $\Upsilon$(4S) resonance with the CLEO III detector.
We use only runs with good lepton identification, which leads to
slightly different, but overlapping, datasets for the electron
and muon modes with integrated luminosities of 6.7 and 8.0 fb$^{-1}$
respectively.

We use the decay chain of $D^{*+} \to D^{0}\pi^{+}$, and $D^{0}$ further 
decays to $\pi \ell \nu$ or $K \ell \nu$. The slow charged pion helps the 
identification and background rejection for $D^{0}$. The experimental 
observable is $\Delta M$ which is the mass difference
of $D^{*+}$ and $D^{0}$.
$D^{0}$ candidates are reconstructed from lepton, hadron ($\pi$
or $K$), and neutrino combinations. A major challenge is the contamination
of the Cabibbo-suppressed $D^{0}\to \pi^{-}\ell^{+}\nu$ sample by 
the Cabibbo-favored $D^{0}\to   K^{-}\ell^{+}\nu$ decays, which are about 
a factor of 10 more common. The use of a Ring Imaging Cherenkov (RICH)
detector and specific ionization in the drift chamber ($dE/dx$) reduces
this contamination dramatically by distinguishing $K$ from $\pi$ mesons.
The resulting efficiency and misidentification probability suppress
misidentified $D^{0}\to   K^{-}\ell^{+}\nu$ decays to 15\% of the
$D^{0}\to \pi^{-}\ell^{+}\nu$ signal. Detailed event selection can be
found in~\cite{Dsemi-cleoiii}.

\begin{figure}
\epsfxsize190pt
\figurebox{190pt}{220pt}{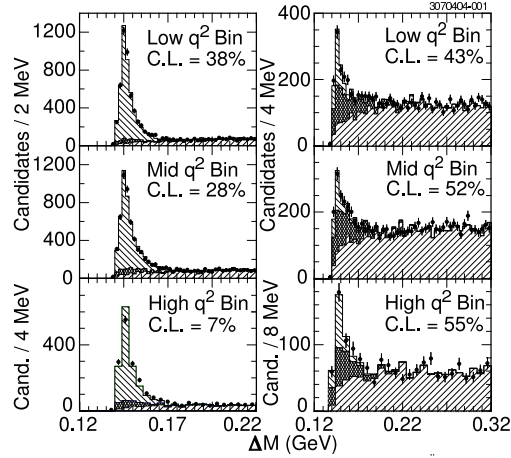}
\caption{The fits to the  $\Delta M$ distributions for 
         $D^{0}\to   K^{-}\ell^{+}\nu$ (left) and 
         $D^{0}\to \pi^{-}\ell^{+}\nu$ (right) and their 
         confidence levels (C.L.). The data (points) are
         superimposed on the sum of the normalized
         simulated signal (peaked histogram), peaking background
         (dark histogram) and false-$\pi_{s}$ background
         (broad histogram).}
\label{3070404-001}
\end{figure}

Figure~\ref{3070404-001} shows the fits to the  $\Delta M$ distributions
for $D^{0}\to   K^{-}\ell^{+}\nu$ and $D^{0}\to \pi^{-}\ell^{+}\nu$ and
their confidence levels (C.L.). We divide the data into three $q^{2}$
bins. The bin size is guided by our $q^{2}$ resolution of 0.4 GeV$^{2}$.
We measure the ratio of the branching fractions,
$R_{0}$ = ${\cal B}(D^{0}\to \pi^{-}\ell^{+}\nu)/
           {\cal B}(D^{0}\to K^{-}\ell^{+}\nu)$ to be
$R_{0}$ = 0.082$\pm$0.006$\pm$0.005. This result is consistent with
the previous world average of 0.101$\pm$0.018~\cite{PDG} but
more precise.
We then use a simple pole parameterization to determine parameters
describing the form factors by fitting the corrected $q^{2}$ distributions.
We find $|f^{\pi}_{+}(0)|^{2}|V_{cd}|^{2}/|f^{K}_{+}(0)|^{2}|V_{cs}|^{2}$
= 0.038$^{+0.006+0.005}_{-0.007-0.003}$. Using $|V_{cd}/V_{cs}|^{2}$
= 0.052$\pm$0.001~\cite{PDG} gives
$|f^{\pi}_{+}(0)|/|f^{K}_{+}(0)|$ = 0.86$\pm$0.07$^{+0.06}_{-0.04} \pm 0.01$.
The details of this work are described in~\cite{Dsemi-cleoiii} and have 
been submitted to PRL.

\section{First CLEO-c results on exclusive $D^{0}$ Semileptonic Decays}

Just like $\Upsilon$(4S) which is an ideal laboratory for $B$ physics 
study at $e^{+}e^{-}$ collider, $\Psi$(3770) offers many advantages for 
Charm physics than $\Upsilon$(4S). The advantages come from the threshold 
production of charm and the unique kinematics constrains associated with 
threshold production. 
The data sample used for this analysis was collected at 
the $\psi(3770)$ resonance with the CLEO-c detector.
It corresponds to an integrated luminosity of 60 pb$^{-1}$. 

\begin{figure}
\epsfxsize190pt
\figurebox{190pt}{240pt}{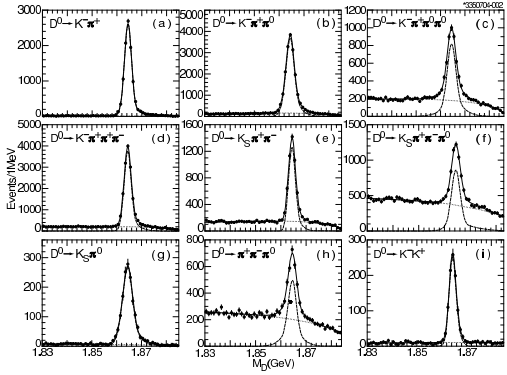}
\caption{Fits to the beam-constrained masses for different 
         fully reconstructed $D^{0}$ decay modes.
         The signal is described by a Gaussian and a bifurcated 
         Gaussian to account for the initial state radiation.
         The background is described by an Argus
         function.}
\label{D0Tag}
\end{figure}

We first select events with a fully reconstructed $D^0$ meson where 
$D^0\to K^-\pi^+$,           $K^-\pi^+\pi^0$, $K^-\pi^+\pi^0\pi^0$, 
$K^-\pi^+\pi^+\pi^-$, $K_S\pi^0$,      $K_S\pi^+\pi^-$, 
$K_S\pi^+\pi^-\pi^0$, $\pi^+\pi^-\pi^0$, and $K^-K^+$. 
Charge conjugate decays are implied. 
Within the tagged events, we select the subset in which 
the $\bar D^0$ meson semileptonically decays to a specific 
final state. The efficiency-corrected ratio of the event yields
gives the absolute branching fraction for the exclusive semileptonic 
decay mode. The selection of the tag $D^0$ candidates is based on two 
variables $\Delta E=E_D-E_{beam}$ 
(the difference between the energy of the tag $D^0$ candidate ($E_D$)  and 
the beam energy ($E_{beam}$) ), and the beam constrained mass 
$M_{bc}=\sqrt{E_{beam}^2-p_D^2}$, where $p_D$ is the momentum of the 
tag $D^0$ candidate.
Fits to the beam-constrained mass distributions for $D^{0}$ 
candidates are shown in Figure~\ref{D0Tag}.

\begin{figure}
\epsfxsize190pt

\centerline{\mbox{\epsfxsize=95pt\epsffile{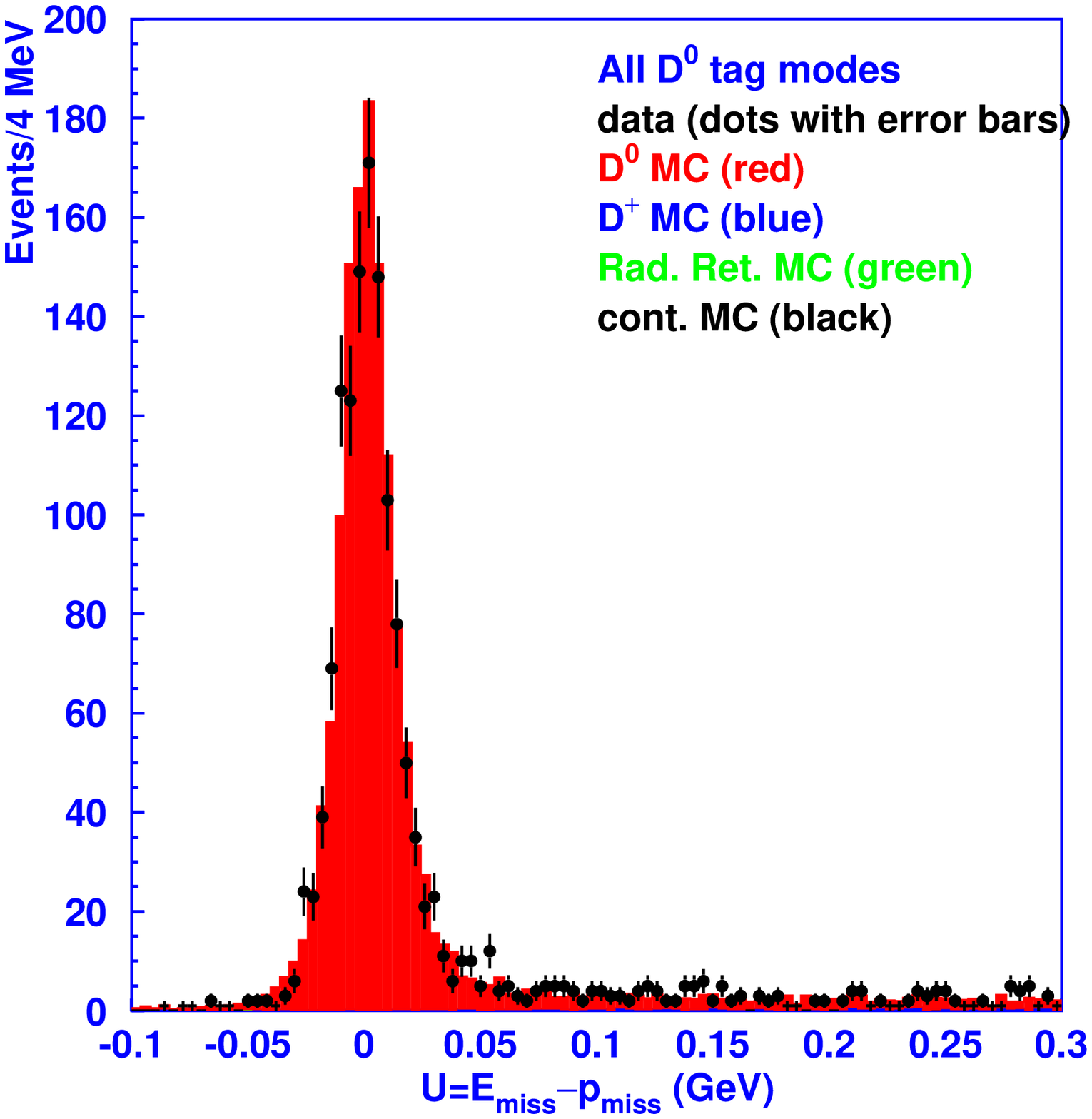}
                  \epsfxsize=95pt\epsffile{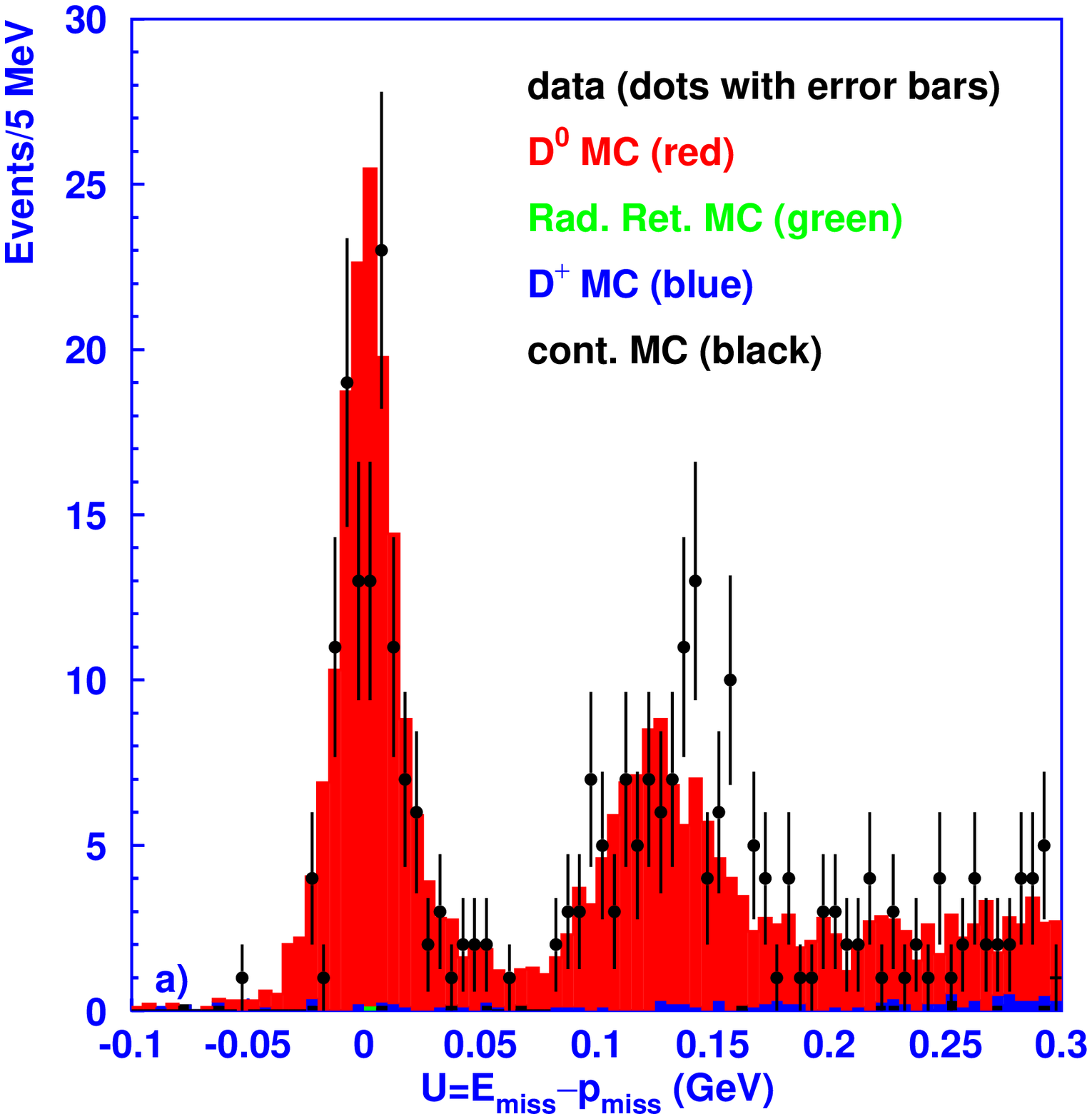}}} 
\centerline{\mbox{\epsfxsize=95pt\epsffile{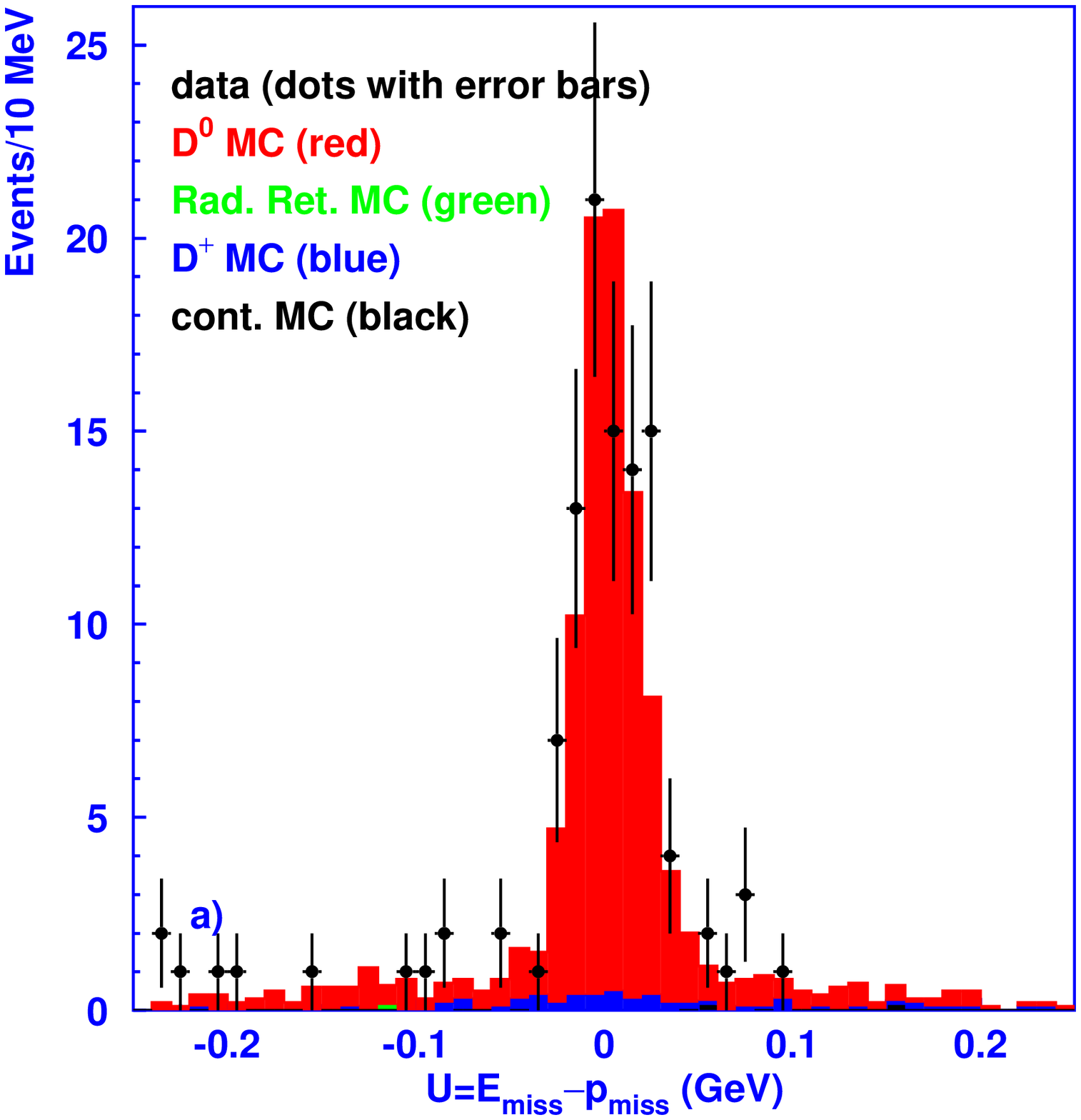}
                  \epsfxsize=95pt\epsffile{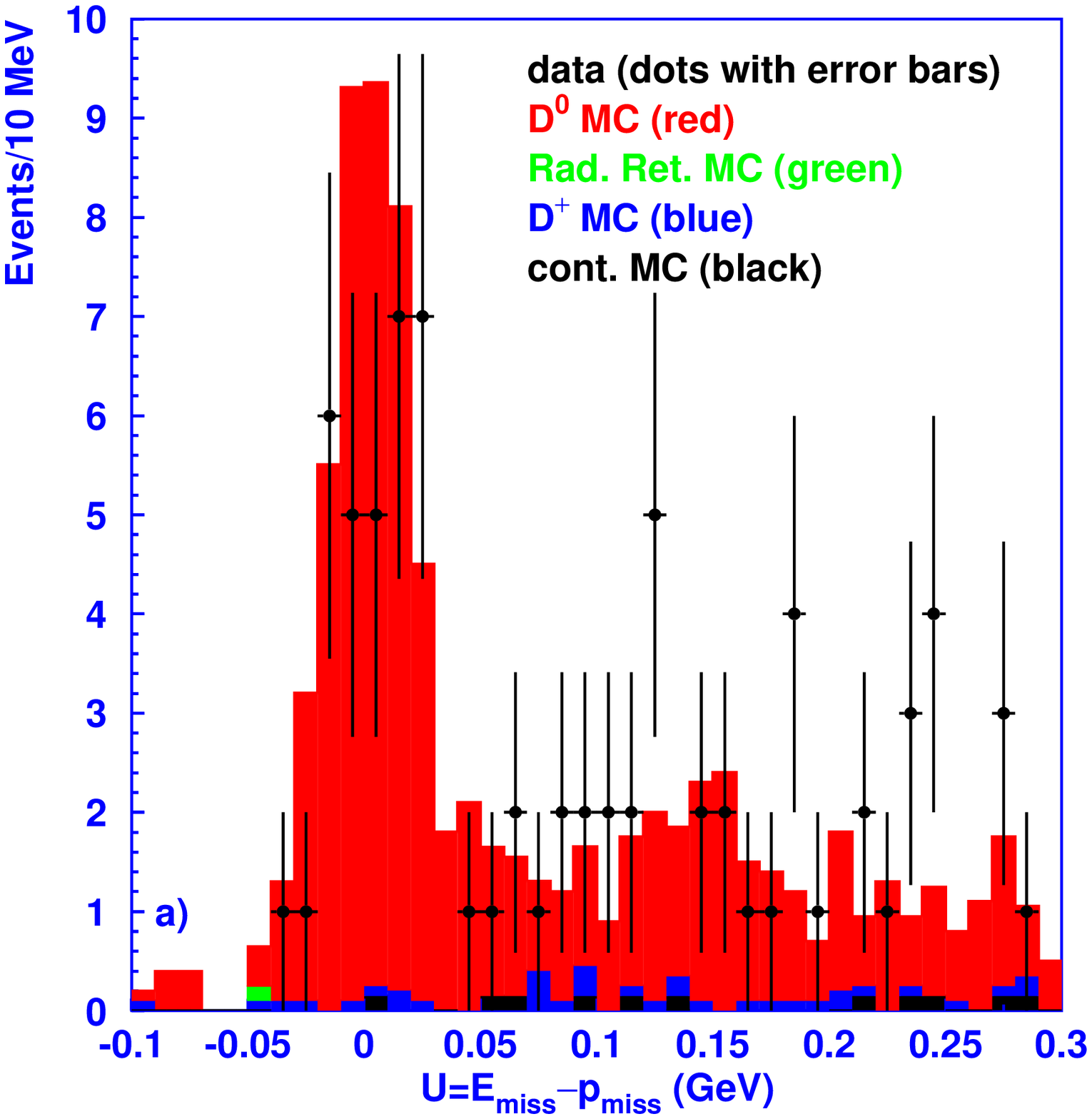}}} 
\caption{$U=E_{\rm miss}-p_{\rm miss}$ for $D^{0} \to K^{-}\ell^{+}\nu$
         (top left), $\pi^{-}\ell^{+}\nu$ (top right), 
         $K^{*-}\ell^{+}\nu$ (bottom left) and 
         $\rho^{-}\ell^{+}\nu$ (bottom right) from data and MC.}
\label{uklnu}
\end{figure}

We then find $D^0$ semileptonic decays into 
$K^{-}\ell^{+}\nu$,  $\pi^{-}\ell^{+}\nu$,
$K^{*-}\ell^{+}\nu$ ($K^{*-}\to K^-\pi^0$), 
and $\rho^{-}\ell^{+}\nu$ ($\rho^-\to\pi^-\pi^0$)
against a tag $D^0$ by reconstructing the difference of the missing 
energy and missing momentum which should peak at zero.  
In Figure~\ref{uklnu}, 
we present the $U$ 
distributions from data and MC for $D^{0}\to   K^{-}\ell^{+}\nu$, 
$\pi^{-}\ell^{+}\nu$, $K^{*-}\ell^{+}\nu$ and
$\rho^{-}\ell^{+}\nu$
The comparison shows good agreement between the data and MC. 

\begin{figure}
\epsfxsize190pt
\figurebox{190pt}{250pt}{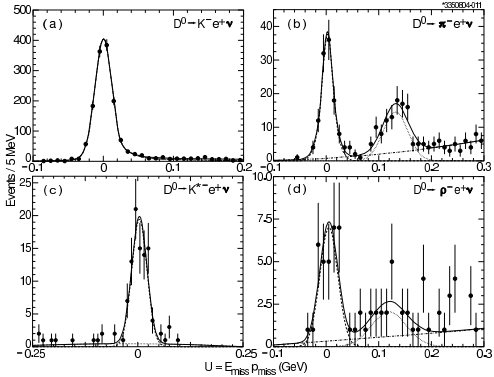}
\caption{Fits to $U=E_{\rm miss}-p_{\rm miss}$ distributions for 
         $D^0 \to K^{-}\ell^{+}\nu$,  $\pi^{-}\ell^{+}\nu$, 
         $K^{*-}\ell^{+}\nu$ and $\rho^{-}\ell^{+}\nu$, with the 
         other $\bar{D^{0}}$ fully reconstructed.}
\label{semi}
\end{figure}

The fits to the $U=E_{miss}-p_{miss}$ distributions 
are shown in Figure~\ref{semi}.
For $D^{0}\to   K^{-}\ell^{+}\nu$, the observed yield is
1405.1$\pm38.5$ which corresponds to an absolute branching
fraction measurement of
${\cal B}(D^{0}\to   K^{-}\ell^{+}\nu)$ = (3.52$\pm0.10\pm0.25$)\%,
comparing with $(3.58\pm0.18)\%$ from PDG.
For $D^{0}\to   \pi^{-}\ell^{+}\nu$, the observed yield is
109.1$\pm$10.9 which corresponds to an absolute branching
fraction measurement of
${\cal B}(D^{0}\to  \pi^{-}\ell^{+}\nu)$ = (0.25$\pm0.03\pm0.02$)\%,
comparing with $(0.36\pm0.06)\%$ from PDG.
For $D^{0}\to   K^{*-}\ell^{+}\nu$, the observed yield is
88.0$\pm$9.7 which corresponds to an absolute branching
fraction measurement of
${\cal B}(D^{0}\to  K^{*-}\ell^{+}\nu)$ = (2.07$\pm0.23\pm0.18)\%$,
comparing with $(2.15\pm0.35)\%$ from PDG.
For $D^{0}\to  \rho^{*-}\ell^{+}\nu$, the observed yield is
30.1$\pm$5.8 which corresponds to an absolute branching
fraction measurement of
${\cal B}(D^{0}\to  \rho^{*-}\ell^{+}\nu)$ = (0.19$\pm0.04\pm0.02)\%$,
this is the first observation and measurement of this decay mode.
We also measure the ratio of absolute branching fractions to be:
${\cal B}(D^{0}\to  \pi^{-}\ell^{+}\nu )\over
 {\cal B}(D^{0}\to    K^{-}\ell^{+}\nu )$ = $(7.0\pm0.7\pm0.3)\%$
($(10.1\pm1.8)\%$ from PDG), and
${\cal B}(D^{0}\to  \rho^{-}\ell^{+}\nu)\over
 {\cal B}(D^{0}\to    K^{*-}\ell^{+}\nu)$ = $(9.2\pm2.0\pm0.8)\%$.

The details of this work can be found in Ref.~\cite{ICHEP04}. 
We can see that most of our measurements based on first 60 pb$^{-1}$ 
of CLEO-c data  are already better than those listed in PDG.
We also present the first observation of $D^{0} \to \rho^{-}\ell^{+}\nu$ 
decay.
The errors are statistical and systematic, respectively. The dominant
systematic error comes from uncertainties of 
track and $\pi^0$ reconstruction efficiency
(3\% per track and 4.4\% per $\pi^0$)
which will improve with a larger data sample and further study. 
Our results for $D^{0} \to K^{-}\ell^{+}\nu$ and 
$D^{0} \to K^{*-}\ell^{+}\nu$ are  consistent with those from the PDG. 
Our result for ${\cal B}(D^{0} \to \pi^{-}\ell^{+}\nu)$ is lower than the PDG value. 
The ratio 
{${\cal B}(D^{0} \to \pi^{-}\ell^{+}\nu)\over{\cal B}(D^{0}\to K^{-}\ell^{+}\nu)$} 
is close to the CLEO III result
$(8.2\pm0.6\pm0.5)\%$~\cite{Dsemi-cleoiii}, while lower than the PDG value.

\end{document}